\documentstyle[12pt,aaspp4,psfig]{article}
\def\etal {{\it et~al.}}
\def\ang   {\AA\/}

\def\msun{M_{\odot}}

\def\mdot{\dot M}
\doublespace
\begin{document}

\title{Simultaneous HST/ASCA 
Observations of LMC X-4:\\ 
X-ray Ionization
Effects on a Stellar Wind}

\author{S. D. Vrtilek\altaffilmark{1}, B. Boroson}   
\affil{Center for Astrophysics, 60 Garden St.,
Cambridge, MA. svrtilek@cfa.harvard.edu, 
bboroson@cfa.harvard.edu} 

\author{F. H. Cheng\altaffilmark{2}} 
\affil{Department of Astronomy and Astrophysics, Villanova University,
Villanova, PA 19085. fcheng@ucis.vill.edu}

\author{R. McCray}
\affil{JILA, Campus Box 440, University of Colorado, Boulder, CO 80309.
dick@jila.colorado.edu}

\and 

\author{F. Nagase} 
\affil{ISAS, 3-1-1, Yoshinodai, Sagamihara, Kanagawa 229, Japan.
nagase@astro.isas.ac.jp}

\altaffiltext{1}{Visiting Professor, Department of Astronomy, University of Maryland, College Park,
MD 20742}
\altaffiltext{2}{Center for Space Science, Shanghai Jaio Tong University,
Shanghai, 200023, People's Republic of China} 

\begin{abstract} 
We present first results from simultaneous
ultraviolet (HST/GHRS) and X-ray (ASCA) observations of the 
13.5s pulsar
LMC X-4 taken in 1996 May.
The ASCA observations covered 1.12 binary orbits (1.58 days)
and the HST observations were centered on this for roughly 0.4 
orbital phase coverage (0.56 days).
The GHRS data are the highest resolution (both temporal and spectral) 
ultraviolet spectra ever taken of
LMC X-4.
With generally-accepted 
parameters for the source, fits to the UV continuum
using a model that incorporates X-ray heating of the companion star 
and the accretion disk yields a mass accretion rate 
$\mdot$~=~4.0~$\times 10^{-8} \msun$~yr$^{-1}$;
the X-ray luminosity implied by this value is  
consistent with the X-ray flux measured during  
simultaneous observations (3.2 $\times 10^{-10}$~ergs~cm$^{-2}$~s$^{-1}$).
The model accurately predicts observed B magnitude and ultraviolet 
variations over both orbital and long-term periods.
The ultraviolet P-Cygni lines show dramatic changes with orbital phase
with strong broad absorption near X-ray eclipse and narrow absorption
when the X-ray source is in the line-of-sight.
We interpret this as a result of X-ray photoionization of the
stellar wind; when the neutron star is in front of the normal star,
the wind absorption disappears and mainly the photospheric absorption lines
are visible.
The X-ray pulse period measured during our observations, 13.5090$\pm$0.0002s,
is consistent with steady spin-down over 
the past 10 years.
No pulsations were detected in the ultraviolet observations with 
upper limits to the pulsed fraction around N\,V and C\,IV of 1.8\% 
and 2.7\% in the 
continuum and 12.4\% and 7\% in the absorption
troughs.
\end{abstract}

\section{Introduction}

LMC X-4 is a 13.5s pulsar 
orbiting a 20$\msun$ O7 III-V companion every 1.4 days
(Kelley \etal~1983; Ilovaisky \etal~1984).
The system exhibits a long term cycle
with a roughly 30-day period  
(Lang \etal~1981; Ilovaisky
\etal~1984) similar to the 35-day period of
Her~X-1. 
Although the high-mass companion of LMC~X-4 implies the presence
of a wind, LMC X-4  has many properties in common with the disk-fed
system Her~X-1.
In both systems the long term periods are roughly 20 times 
the orbital periods.
Both systems are fully eclipsing 
(Tananbaum \etal~1972; Li \etal~1978); show 
a power-law spectrum with a soft X-ray ``excess"
in the 0.5-10 keV band (Dennerl 1989);
optical light curves 
with a power spectrum that
shows power at the sum of the orbital and long term
frequencies, but not in the difference,
implying that the long-term variations are due to precession
of an accretion disk (Ilovaisky \etal~1984);
and correlation
between times of low hard X-ray luminosity and episodes of spindown
(Dennerl 1991; Wilson \etal~1994).
The optical lightcurves of
LMC X-4 show changes in peak-to-peak amplitude of up to 40\% relative 
to the mean over the 30-day
period whereas the optical lightcurves of Her X-1 show less than 20\%
change in peak-to-peak amplitude over the 35-day cycle.
For LMC X-4 the X-ray flux varies by a factor of 60 between the high
and low states of the 30-day period;  however, the flux of scattered
X-rays, measured during X-ray eclipses, remains unchanged throughout
the 30 day cycle (Woo \etal~1995), implying that the low state is 
not caused by a decrease in X-ray luminosity but by attenuation
in intervening matter.
LMC X-4 also displays
flares which typically occur once a day 
(Dennerl 1989; Levine \etal~1991; Woo \etal~1996).
The X-ray luminosity during the flares rises
to $\sim$5 times the Eddington
luminosity and the spectrum changes from a power law to a very hot
thermal spectrum.

There are several mechanisms that could cause the UV continuum or lines to
vary with the 13.5 second neutron star rotation period.
In Vela~X-1 it has been shown
that the Si\,IV and N\,V P~Cygni lines of Vela~X-1 vary  
with the 283 second pulsar period, presumably as a result of
time-dependent 
photoionization of the stellar wind by the X-rays
(Boroson \etal~1996a). The UV
continuum could pulsate as a result of X-ray heating of either the normal
star or the accretion disk, as is the case in Her~X-1 (Boroson
\etal~1996b).

Here we present coordinated ultraviolet and X-ray observations of 
LMC X-4 taken with the GHRS on HST and with the GIS and SIS on ASCA.  
The high temporal resolution of the GHRS allowed us to search for
a manifestation of the pulsations in the UV and the high spectral 
resolution allows us to study
the geometry of the system as reflected in the line profiles.
Some results from an ASCA observation covering a full binary orbit
taken an year earlier are also presented.
We compare the HST data as well as archival IUE data with models
predicting UV continuum emission from the X-ray heated disk and star
that has been successfullly applied to the Her X-1 system.
We interpret the dramatic changes with orbital phase observed in 
the ultraviolet spectra in terms of  
the effects of X-ray photonionization on the stellar wind
of the normal companion.  The observations and analysis are described
in section 2 and our interpretation 
is discussed in section 3.

\section{Observations and Analysis}

Figure 1 shows the location of our observation in comparison with
the one-day averaged lightcurves obtained from the All Sky Monitor on
XTE  
and the same data folded with the long term ephemeris and period of 
Dennerl \etal~(1992). 
This indicates that our simultaneous HST and ASCA observations 
occurred during the 
high state of the 30-day cycle,
corresponding to a phase coverage of $\phi_{30-day}$=0.20-0.22.
 
The ASCA lightcurves and HST coverage are plotted on Figure 2.
HST took eight seperate observations, but as the target was 
nearly in HST's Continuous
Viewing Zone the gaps between the observations are small.
A journal of the observations is given in Table 1.
HST captured an eclipse 
egress that did not have simultaneous X-ray coverage.
The HST observations were taken with the GHRS G160M grating
centered alternately at 1240 \AA~(N V) and 1550 \AA~(C IV) in the RAPID
mode with a time resolution of 0.5s.
Our use of the GHRS
RAPID mode prevented us from over-sampling the spectrum, checking for bad 
counts, and correcting for the Doppler shift due to the spacecraft orbit.
The observed flux in the  N\thinspace V $\lambda 1240$ line measured with IUE
(van der Klis \etal~1982) ranges from
$F \approx 1 \times 10^{-12}$ergs cm$^{-2}$
$s^{-1}$
at minimum ($\phi_{orb} =  \approx 0.5$) to
$F \approx 3 \times 10^{-12}$ ergs cm$^{-2}$ s$^{-1}$
at maximum ($\phi_{orb} =  \approx 0.9)$.
The fluxes measured from our observations are listed in Table 1.

The 1996 ASCA observations caught an eclipse ingress and
two pre-eclipse dips that
did not have simultaneous UV coverage.
The 1994 ASCA observations covered more than a full binary orbit including
an eclipse.
Detailed studies of the ASCA observations will appear elsewhere 
(Boroson \etal~1997b).
No flares occurred during either the 1994 or 1996 ASCA observations 
although each covered more than a
binary orbit and flares are reported to occur roughly once a day.
It is possible that flares (which last roughly 30 minutes) occurred during
the data gaps (roughly 40-50 minutes) due to earth occultation.

\subsection{Continuum Fits}
A model involving X-ray heating of the disk and star as previously applied
to Her X-1 (Cheng, Vrtilek, \& Raymond~1995) was adapted for use with LMC X-4.
Changes to the model from that described in Cheng, Vrtilek, \& Raymond
are inclusion of both gravity darkening and limb darkening effects 
(important for the massive companion in LMC X-4) as well as a different
reddening curve towards the LMC (Nandy~\etal~1981).   
We use a distance to the source of 50 kpc 
and an E$_{B-V}$ of 0.05.
The companion star is modelled as a 20 $\msun$ O7III star with an 
effective temperature of 35,000K 
and IUE spectra of stars of varying temperature are used to determine 
the spectral shape at different points on the star and disk surface. 
Predictions of the model from the ultraviolet
and optical B band are shown in Figure 3.  We are using archival
IUE data for comparison, hence 
we restrict ourselves to the
region near C~IV since the presence of strong geocoronal Lyman Alpha
prevents accurate determination of the IUE flux near N~V.  
In addition our model, which is constructed to predict continuum emission, 
consistently overpredicts the flux near N~V in this system; this is 
likely due to the presence
of a stellar wind---not included in the model---which results in 
strong absorption at N~V. 
Because the 30.25d period has a large uncertainty the long term phase of the
IUE observations cannot be determined. 
The ultraviolet data including
the current observations and all archival IUE observations fall within extremes
given by a lower limit to $\mdot$ of 
3.2~$\times~10^{-9}\msun$ yr$^{-1}$ and an upper limit of 
1.0~$\times~10^{-7}\msun$ yr$^{-1}$.
The best fit curve to the HST continuum data (shown as filled triangles) 
corresponds  
to an $\mdot$ of 4.0~$\times~$10$^{-8}\msun$ yr$^{-1}$. 
There is uncertainty in the flux level, both for the model and the data. 
Our HST continuum
determination is based on a very narrow wavelength band 
since the observations were centered on lines and the total 
available bandpass was
only 37$\ang$ for the spectral resolution we required; 
the HST data are
absolute flux calibrated to only 10\%; there is a difference in 
resolution between IUE and GHRS.  As for the model, the uncertainties in the
reddening and stellar temperature can shift the flux. The fluxes
near C IV and N V show a similar flux vs. orbit trend.  The relative
errors in flux are likely to be much smaller than the overall flux
normalization.  

The change in peak-to-peak amplitude of the optical flux from the 30-day 
low to the 30 day 
high 
predicted by the model is 
consistent with that observed (Ilovaisky \etal~1984).  
Figure 3b shows the extremes over the 30.25 day cycle for B magnitude 
variations and Figure 3c shows
the model predictions at the X-ray on and off states for
a given $\mdot$ reproducing the variations shown by 
Ilovaisky~\etal~1984 in their Figure 1. 
During binary eclipse the total UV and optical flux is due to the 
companion star.  During the time of maximum UV flux (binary phase 0.75)
for the highest $\mdot$ the contribution from the disk is 8\% and from
the heated star 10\%.
A comprehensive presentation of the fits to individual IUE spectra and 
archival optical data will appear in a later paper 
(Preciado \etal~1997).

The X-ray spectra obtained by ASCA are consistent with a power law
of photon index $\alpha$ = 0.63$\pm$0.02, a 0.180$\pm$.006 keV blackbody, 
and an iron line at
6.54$\pm$0.03 keV with an equivalent width of 99$\pm$25eV, typical
for this source over the energy range observed.
The 2-10 keV flux is 
2.9~$\times 10^{-10}$ergs~s$^{-1}$~cm$^{-2}$ with about 10\% changes from
the mean during the period of observations. 
The luminosity obtained from the $\mdot$ measured with the UV continuum
fits, using L$_x$~=~0.5GM$\mdot$/r, is consistent with the luminosity implied
by this flux.

\subsection{Spectral Features} 

Figure 4 shows the N~V and C~IV line profiles at four orbital phases each;
these are different for the two lines since they could not be observed
simultaneously at this spectral resolution.
These two lines
are the strongest in the UV spectrum and
are known from IUE observations
to be strongly and weakly phase dependent, respectively.
The C~IV profiles show little evolution from binary phase 0.15-0.49.
At binary phase 0.11, the observation closest to the X-ray eclipse
(binary phase 0.0), we see
broad absorption in the N~V doublets.  The maximum velocity at this
phase (1150 km/s) is reasonable for a terminal velocity for the wind
however it is likely that by phase 0.11 some ionization has already
taken place and terminal velocities of up to 1300 km/s are possible
(Boroson \etal~1997a).  As we progress in phase
towards seeing more of the X-ray source we find that the broad absorption
disappears and what remains are relatively narrow absorption profiles.
The width of the residual absorption left at phase 0.41 in N~V is
consistent with the width of the optical lines which Hutchings \etal~
attributed to the photosphere. 
Comparison with low-resolution IUE archival data suggests that the
variations are phase dependent effects, presumably arising in the 
geometry-dependent interaction between the X-ray source and stellar wind.

Interstellar absorption lines from S\,II$~\lambda\lambda1250.578,1253.805$
are visible in our exposures in the wavelength region near 
N\,V$~\lambda\lambda1238,1242$.  The velocities of these absorption
lines (heliocentric redshifts are consistent with zero within the errors
of our measurement) suggest that they arise in gas within the Galaxy, and 
not in the
LMC (where we would expect velocities of the order 280-320km~s$^{-1}$;
Bomans \etal~1996).  
The narrow C\,IV absorption lines are superimposed on broader absorption
features, which we interpret as arising in the photosphere of the O~star in
the LMC~X-4 system. As a result of the overlap, the velocity and
especially the equivalent widths of the interstellar C\,IV lines are
uncertain.  The heliocentric redshifts of 
the C\,IV lines are also consistent with zero within the uncertainties of
our measurement implying that the  
C\,IV absorber is also within the Galaxy.  

There is no evidence for a C\,IV absorber associated with LMC~X-4, which
might be expected if the X-ray source photoionizes surrounding
interstellar gas (McCray, Wright, and Hatchett 1977).  The N\,V line at
$\phi=0.41$ shows what appears to be a narrow feature at the base of the
absorption due to the blue doublet component.  However, observations at
higher signal-to-noise ratio are needed to establish that this results from an
interstellar feature and not merely a fluctuation in the photospheric line
profile. 

\subsection{Search for Pulsations}

The pulse periods that we measured from ASCA data of 1994
(13.5069$\pm$0.0002s) and 1996  
(13.5090$\pm$0.0002s)
are consistent with spin-down over the past 10 years (Figure 5).
While earlier observations are scant they indicate that 
sometime between 10 and 20 years ago LMC X-4 went through a period
of spin-up.
Such rapid, erratic, changes from spin-up to spin-down are expected
from equilibrium rotators (Chakrabarty \etal~1997), systems in which
the spin period equals the Kepler period near the inner boundary of
the accretion disk.

We searched for pulsations in the UV spectrum at the 13.5 second
pulsar period using 
an analysis of variance (ANOVA) method
(Davies 1990, 1991).  This approach, which involves binning the data
over trial periods, is well-suited to data with gaps or with non-uniform
readout times.  To detect pulsations from the accretion disk, we need to  
subtract the changes in light-travel time from the orbiting pulsar to the
orbiting HST from the 
uniform readout times of the GHRS.  Thus we report only the ANOVA method 
and not a power-spectral search to find pulsations from material 
moving with the neutron star.

From the 4 HST orbits we find a 5$\sigma$
limit of 1.8\% for the fractional peak-to-peak pulse amplitude in the continuum 
centered at 1240\AA~
surrounding the N\,V line.  For the continuum surrounding C\,IV,
the limit is 2.7\%.
Our limits for pulsation 
in the P~Cygni absorption troughs are 12.4\% in the N\,V line and 
7\% in the C\,IV line.

\section{Discussion and Conclusions}

A model that incorporates X-ray heating of the companion star
and the accretion disk 
provides good fits to the continuum UV emission from LMC X-4.
The value of $\mdot$ derived from GHRS observations is consistent 
with that from X-ray flux measured during
simultaneous observations.  
Owing to the size and temperature of the companion the major
contribution to the ultraviolet and
optical flux is the unheated companion star.  
At maximum light (binary phases 0.25 and 0.75) the  
contribution of the unheated star is 82\%;
heating of the primary contributes 10\% and the
disk 8\%.

Although we captured no flares during our observations
the X-ray heating model can accomodate flares by using unsteady accretion
from the stellar wind.  During flares the intensity can increase 
by factors of up to 20 for times
ranging from $\sim$20s to 45 minutes, resulting in super Eddington 
luminosities (Dennerl 1989; Woo \etal~1995). The necessary $\mdot$ is 
3.2~$\times~10^{-7}~\msun$~yr$^{-1}$ 
and implies a range of a factor of 100 in $\mdot$ for LMC X-4.
By contrast the flaring states of Z-source LMXBs such as Sco X-1 and Cyg X-2,
 where the
flaring state is also associated with super-Eddington accretion, 
are produced with changes of only a factor of 2-4 in $\mdot$; this is because
Sco X-1 and Cyg X-2 are disk fed systems normally accreting
at just under the Eddington limit.

It is possible that the greater range and overall greater $\mdot$ required for 
LMC X-4 is due to some intrinsic difference between   
X-ray binaries in the Magellanic Clouds
compared with systems within our own
Galaxy.  The mean luminosity of those sources in the Clouds which have massive
OB-type companions similar to LMC X-4 is 
50 times that of the counterparts
in our Galaxy (van Paradijs \& McClintock 1995).  It has been
suggested that this is due to the lower abundances of metals in the
Clouds.  One linking mechanism
is the effect of X-ray heating of gas as it
falls toward a compact object:  this depends strongly on the atomic
number Z via the photoelectric cross section ($\sigma \propto Z^4$).
For spherical accretion, such heating can seriously impede the
accretion flow and thereby reduce the limiting luminosity to a value
far below the Eddington limit; so the LMC sources may be more luminous because
their low-Z accretion flow is less impeded by heating.  
A second metallicity dependent effect
takes place in wind-fed systems.  The accretion rate depends sensitively
on the velocity of the stellar wind; all available evidence indicates that
the terminal wind velocity decreases with Z.  Such behavior is predicted
by successful theories of radiation-driven stellar winds (Kudrizki \& 
Hummer 1990; Kudritzki~\etal~1991).
In this case the lower metallicity in the LMC means a lower terminal velocity, 
a higher $\mdot$ and a more luminous X-ray source. 

The dramatic orbital variations shown by the N~V profiles can be 
interpreted in terms of X-ray photoionization of the stellar wind
from the companion. 
The narrow absorption lines shown in Figure~4d 
can be attributed to the
surface of the companion star 
and the broad lines to
the wind.  In this scenario,  
the X-ray source
ionizes nearly the entire stellar wind that is not in the shadow of
the companion star, so that 
when the neutron star is in front of the normal star,
the wind absorption disappears and mainly the photospheric absorption lines
are visible.
It is only near phose 0.0, i.e. during X-ray eclipse, that the broad 
lines that reveal the high wind velocities become visible.
Detailed fits to the line profiles with a model that takes into account
the structure of the wind
and its influence on spectral features will be presented in a later paper
(Boroson \etal~1997a).

The pulse amplitude in the X-rays is lower in LMC X-4 
(the pulsed fraction of LMC X-4 is 10\% in quiescence) 
than in Her~X-1 and Vela~X-1.  
In Her~X-1 the X-ray pulsed fraction is close
to 50\% and
UV~continuum pulsations have an amplitude of $\sim0.5$\% (Boroson \etal~1996).
If LMC X-4 resembles Her X-1 than we would expect UV~continuum pulsations
with an amplitude of $\sim0.1$\%, which is below our detection limit.
The narrow wavelength regions surrounding the
lines provided too low a count rate to detect
pulsations similar to those seen in other systems.
In Vela X-1 the pulsed fraction
for 2-10 keV X-ray is 0.3, and, while no continuum pulsations were detected in
the UV using FOS observations, 3\% pulsations in the P-Cygni lines were 
detected.
P~Cygni line pulsations, 
seen in Vela~X-1, may be smeared by light travel time in LMC~X-4,
owing to a more extended 
X-ray photoionized region.   

The STIS, recently installed on HST, extends in several significant
ways the 
ultraviolet capabilities that became available with the GHRS:
with the echelle grating it is possible to sample continuously a broad
region (600$\ang$) of the spectrum at greater spectral
resolution than with the GHRS.  Observations with the STIS that cover binary
phase 0.0 can confirm our interpretation of the line profiles, provide
improved measures of the continuum, 
and enable a more sensitive search for UV manifestation of the 
X-ray pulses.

Until recently, long-term approximately-periodic variability attributed 
to disk precession
has been
known for only three systems:  Her X-1, LMC X-4, and SS433.  XTE
observations suggest similar long-term behavior in SMC X-1 and
Cyg X-2 (Levine \etal~1996; Wijnands, Kuulkers, \& Smale 1996).
The interpretation of these periods in terms of disk precession has been
questioned
by several authors including, e.g., Kondo \etal~1983, who showed that
precession of a disk controlled by the gravitational fields of the
neutron and companion stars is untenable:  any induced precession disppears
rapidly because of differential precession within the disk.
Recently Iping and Petterson (1990) proposed that the
behavior attributed to precession is maintained by
the influence of the X-ray emission on the structure of the disk.
Accordingly, in these systems, the tendency of disks to undergo periodic 
changes in their orbital  
orientations is more appropriately termed radiation-driven warping. 
Iping and Petterson's primary result, that strong central illumination can
maintain disk warping, has been obtained
analytically by Pringle (1996) and Maloney, Begelman, \& Pringle (1996). 
The success of our X-ray heated star plus warped disk model in 
fitting the optical, UV, and X-ray lightcurves for both the orbital
and long-term periods of Her X-1 and LMC X-4 
supports an interpretation in terms of radiation-driven warping. 
\vskip 0.2in
Based on observations with the NASA/ESA {\it Hubble Space Telescope},
obtained at the Space Telescope Science Institute, which is operated
by the Association of Universities for Research in Astronomy, Inc.,
under NASA contract GO-05874.01-94A.
We are grateful for quick-look results provided by the ASM/RXTE team.
SDV and BB were supported in part by NASA 
(NAG5-2532, NAGW-2685), and NSF 
(DGE-9350074).
\newpage
\normalsize
\centerline{\bf References}
\vskip 0.1in
\baselineskip=15pt
\parindent=-20pt

Bomans, D.J., DeBoer, K.S., Koornneef, J, \& Grebel, E.K. . 1996, A\&A 
313, 101

Boroson, B., McCray, R., Kallman, T.R., \& Nagase, F. 1994, ApJ,
465, 940.

Boroson, B., McCray, R., Kallman, T., \& Nagase, F. 1996a,
ApJ, 465, 940.

Boroson, B., Vrtilek, S.D., McCray, R., Kallman, T., \& Nagase, F.
1996b, ApJ, 473 1079. 

Boroson, B., Vrtilek, S.D., Kallman, T., \& McCray, R. 1997a, in
preparation.

Boroson, B., Vrtilek, S.D., Xu, C., Kelley, R., \& Stahle, C. 1997b, in
preparation.

Chakrabarty, D. \etal~1997, ApJ, 474, 414.  

Cheng, F.H., Vrtilek, S.D., \& Raymond, J.C. 1995, ApJ, 452, 825.

Davies, S.R.  1990, MNRAS, 244, 93.

Davies, S.R.  1991, MNRAS, 251, 64p.

Dennerl, K. 1989, PhD thesis at Max Planck Institute for...(D89).

Dennerl, K. 1989, in 23rd ESLAB Symp., ed. J. Hunt
\& B. Battrick (Noordwijk:ESA-ESTEC), 39.

Dennerl, K. \etal~1992, in Lecture Notes in Physics, 416, New
Aspects of Magellanic Cloud Research, ed. B. Baschek, G. Klare, 
\& J. Lequeux (Berlin: Springer), 74.

Heemskerk, M.H.M., \& van Paradijs, J. 1989, A\&A, 223, 154.

Ilovaisky, S.A., Chevalier, C., Motch, C., Pakull, M., Van Paradijs, J., 
\& Lub, J.~1984, A\&A, 140, 251.

Iping, R.C., \& Petterson, J.A. 1990, A\&A, 239, 221.

Kelley, R.L., Jernigan, J.G., Levine, A., Petro, L.D., \& 
Rappaport, S.~1983, ApJ, 264, 568.

Kondo, Y., Van Flandern, T.C., \& Wolff, C.L. 1983, ApJ, 273, 716.

Lang, F. L., \etal~1981, ApJ, 246, L21.

Levine, A. M., Bradt, H., Cui, W., Jernigan, J. G., Morgan, E. H., 
Remillard, R., Shirey, R. E., \& Smith, D. A. 1996, ApJ, 469, L33.

Levine, A., Rappaport, S., Putney, A., Corbet, R., \& Nagase, F.
1991, ApJ, 381, 101.

Li, F. \etal~1978, Nature, 271, 38.

Maloney, P. R., Begelman, M. C., \& Pringle, J. E. 1996, ApJ, 472, 582.

McCray, R., Wright, C., and Hatchett, S.  1977, ApJ, 211, L29.

Nandy, K, Morgan, D.H., Willis, A.J., Wilson, R., \& Gondhalekar, P.M. 
1981, MNRAS, 196, 955.

Preciado, M., Boroson, B., Vrtilek, S.D., \& Raymond, J.C. 1997,
in preparation.

Pringle, J.E. 1996, MNRAS, 281, 357.

Tananbaum, H., et al. 1972, ApJ 174, L143.

van der Klis, M. \etal~1982, A\&A, 106, 339.

van Paradijs, J. \& McClintock, J.E., 1995, in X-ray Binaries,
eds. W.H.G. lewin, J. van Paradijs, \& E.P.J. van den Heuvel,
Cambridge University Press: Cambridge.

Wijnands, R. A. D., Kuulkers, E., \& Smale, A. P. 1996, ApJ, 473, L45.

Wilson, R.B., Finger, M.H., Pendleton, G.N., Brigg, M., \& 
Bildsten, L. 1994, in The Evolution of X-ray Binaries, eds.
S.S. Holt, \& C. S. Day, AIP Press: New York.

Woo, J.W., Clark, G.W., \& Levine, A.M. 
1995, ApJ, 449, 880.

Woo, J.W., Clark, G.W., Levine, A.M., Corbet, R.H., \& 
Nagase, F. 1996 ApJ 467, 811
\newpage
\centerline{\bf Figure Captions}
\vskip 0.1in
{\bf Figure 1.} (a) One day averages of the flux observed from 
LMC X-4 with the All Sky Monitor 
on board the ROSSI X-ray Timing
Explorer.  (Quick-look results were provided by the ASM/RXTE team.) 
The arrow indicates the time of the simultaneous HST/ASCA observations,
and the horizontal bar represents one 30.25 day interval.
(b) The data from (a) (light lines)
with the lightcurve from (c) (dark lines) superposed.
(c) The data from (a) folded with the ephemeris for the long-term
period provided by Dennerl \etal~1992 ( P$_{30}$ = 30.25+/-.03d and
$\phi^0_{30}$ = JD 2,448,226.0).  Arrow indicates the phase
 of the
simultaneous HST/ASCA observations.
The errors are 1$\sigma$ from counting statistics.
\vskip 0.1in
{\bf Figure 2.} 
(a) The exposure times
of the HST observations which covered the binary phases
0.08-0.49 (start time = JD 2,450,228.44).  The first observation
is N~V and is alternated with C~IV.
(b) ASCA lightcurves (0.5-10 keV):  observations 
covered the binary phases 0.75-1.84 (start time =
JD 2,450,227.92) determined from the ephemeris of
Woo \etal~(1996) with P$_{orb}$ = 1.40840249$\pm6.0\times10^{-7}$d and
$\phi^0_{orb}$ = JD 2,446,729.84878$\pm$.0041d.  
\vskip 0.1in
{\bf Figure 3.} (a) Average UV continuum flux near C~IV {\it vs.} binary phase.
The solid squares represent the GHRS observations.  The open triangles 
are archival IUE data.  The dashed line represents a 30.25 day phase when
the X-ray flux is in the low state; the solid line a 30.25d phase during
an X-ray high state; and the  
dotted line represents the best
fit continuum for the GHRS observations. 
(b) The B magnitude predictions of the model for the three states depicted
in 3a.  (c) The B magnitude 
predictions for a given $\mdot$ during the 
X-ray on and off states. (b) and (c) should be compared with 
Figure 1 in Ilovaisky~\etal~1984. 
\vskip 0.1in
{\bf Figure 4.} The GHRS observations of N~V and C~IV.
\vskip 0.1in
{\bf Figure 5.} The pulse period history of LMC X-4.
\newpage
\centerline{\bf Table 1: Log of GHRS Observations}
\small
\begin{center}
\begin{tabular}{cccccc}\hline
HST&&Orbital&&Wavelength&Continuum\\
Observation & Start time & phase$^1$ & Duration & range&Flux\\
Number&(JD-2,450,000)&at midpoint&(s)&(\AA)&
(ergs~cm$^{-2}~$s$^{-1}~\ang^{-1}$)\\
\hline
\hline
Z3AA0104T & 228.4428 & 0.111 & 4352   & 1222.5-1258.6 &
4.364$\times 10^{-13}\pm$0.008$^{2}$\\
Z3AA0106T & 228.5083 & 0.159 & 4704  & 1532.3-1567.5&
2.673$\times 10^{-13}\pm$0.004$^{3}$\\
Z3AA0108T & 228.5798 & 0.209 & 4484  & 1225.5-1258.6&
4.564$\times 10^{-13}\pm$0.008$^{2}$\\
Z3AA010AT & 228.6511 & 0.259 & 4424  & 1532.3-1567.5&
2.761$\times 10^{-13}\pm$0.004$^{3}$\\
Z3AA010CT & 228.7223 & 0.312 & 4423  & 1225.5-1258.6&
4.457$\times 10^{-13}\pm$0.008$^{2}$\\
Z3AA010ET & 228.7926 & 0.360 & 4543  & 1532.3-1567.5&
2.600$\times 10^{-13}\pm$0.004$^{3}$\\
Z3AA010GT & 228.8619 & 0.410 & 4721  & 1225.5-1258.6&
4.177$\times 10^{-13}\pm$0.007$^{2}$\\
Z3AA010IT & 228.9285 & 0.465 & 6628  & 1532.3-1567.5&
2.411$\times 10^{-13}\pm$0.004$^{3}$\\
\hline
\end{tabular}
\end{center}
$^1$Using the ephemeris of
Woo \etal~(1996) with the nth eclipse in JD given by
a0~+~a1*n~+~a2*n$^2$
where a0~=~JD~2,446,729.84878$\pm$0.0041,
a1~=~1.40840249$\pm~6.0~\times~10^{-7}$ days, and
a2~=~-1.45E-9. $\phi^0_{orb}$~=~a0 and $P_{orb}$~=~a1~+~a2*n.\\
$^2$Flux near N~V ([1226-1234$\ang$]+[1255-1259$\ang$]).\\
$^3$Flux near C~IV ([1532-1544$\ang$]+[1556-1567$\ang$]).\\
\vskip 0.2in

\vfill\eject

\newpage
\section{FIGURES}

\centerline{\psfig{figure=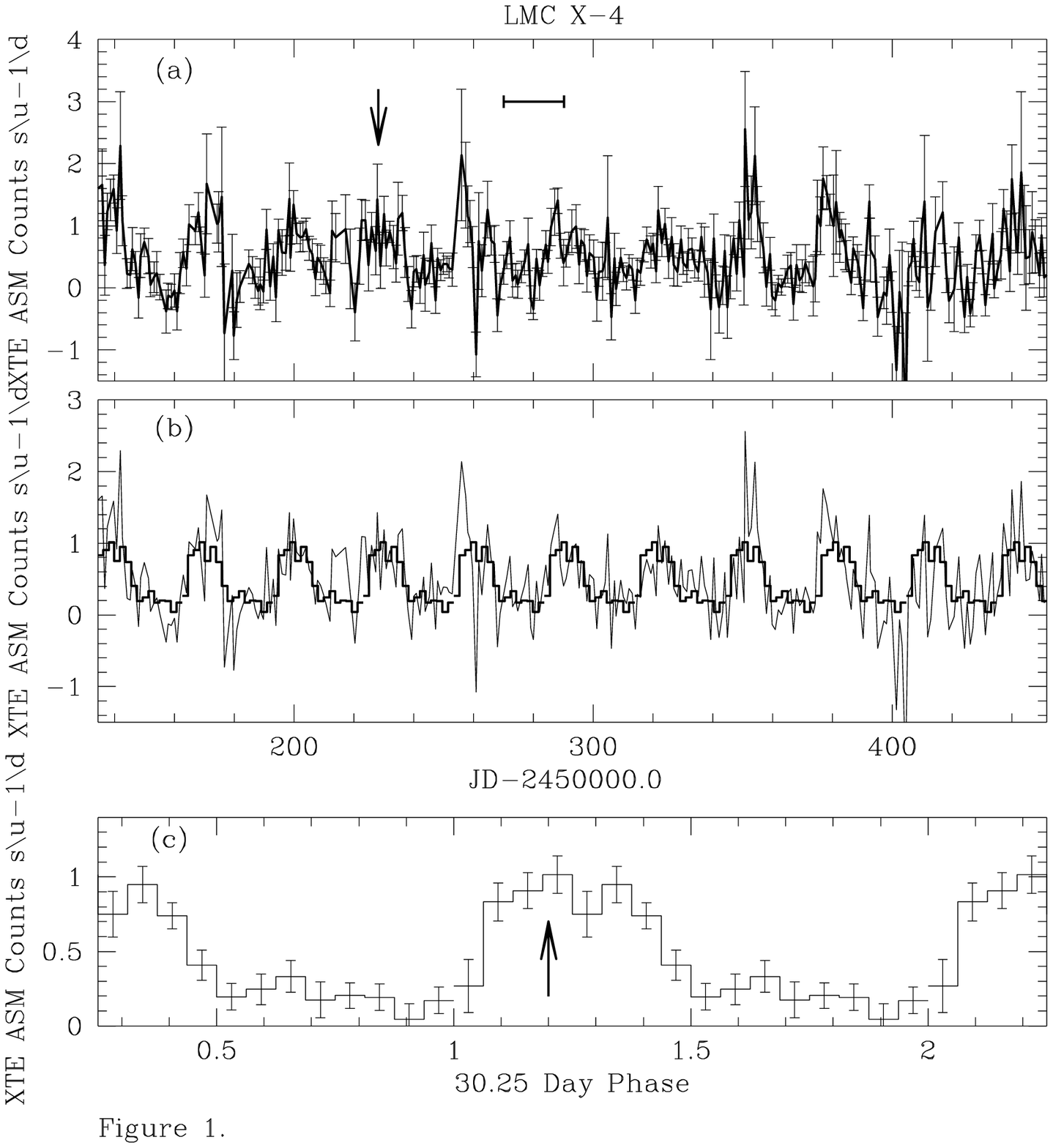}}

\newpage

\centerline{\psfig{figure=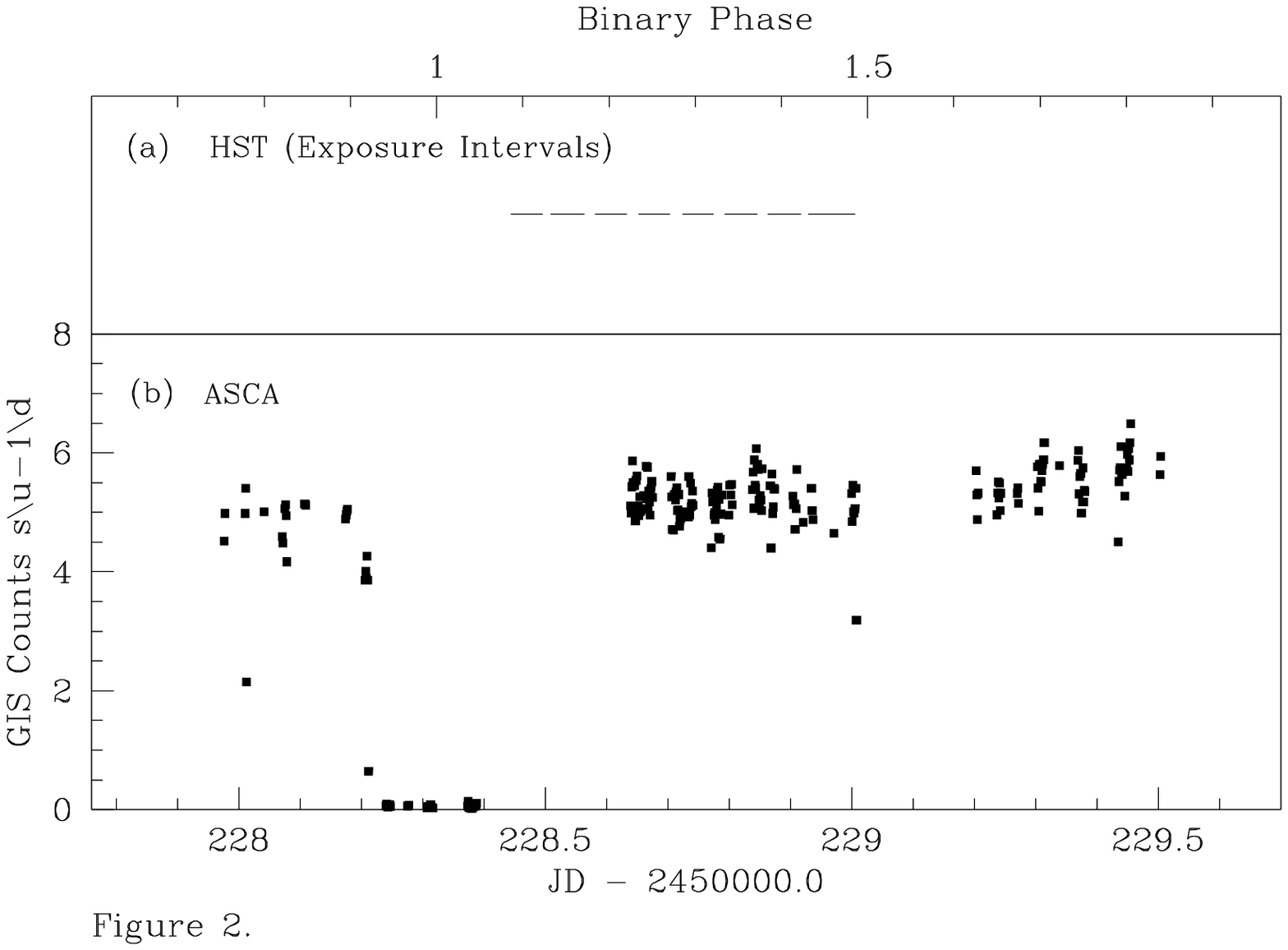}}

\newpage

\centerline{\psfig{figure=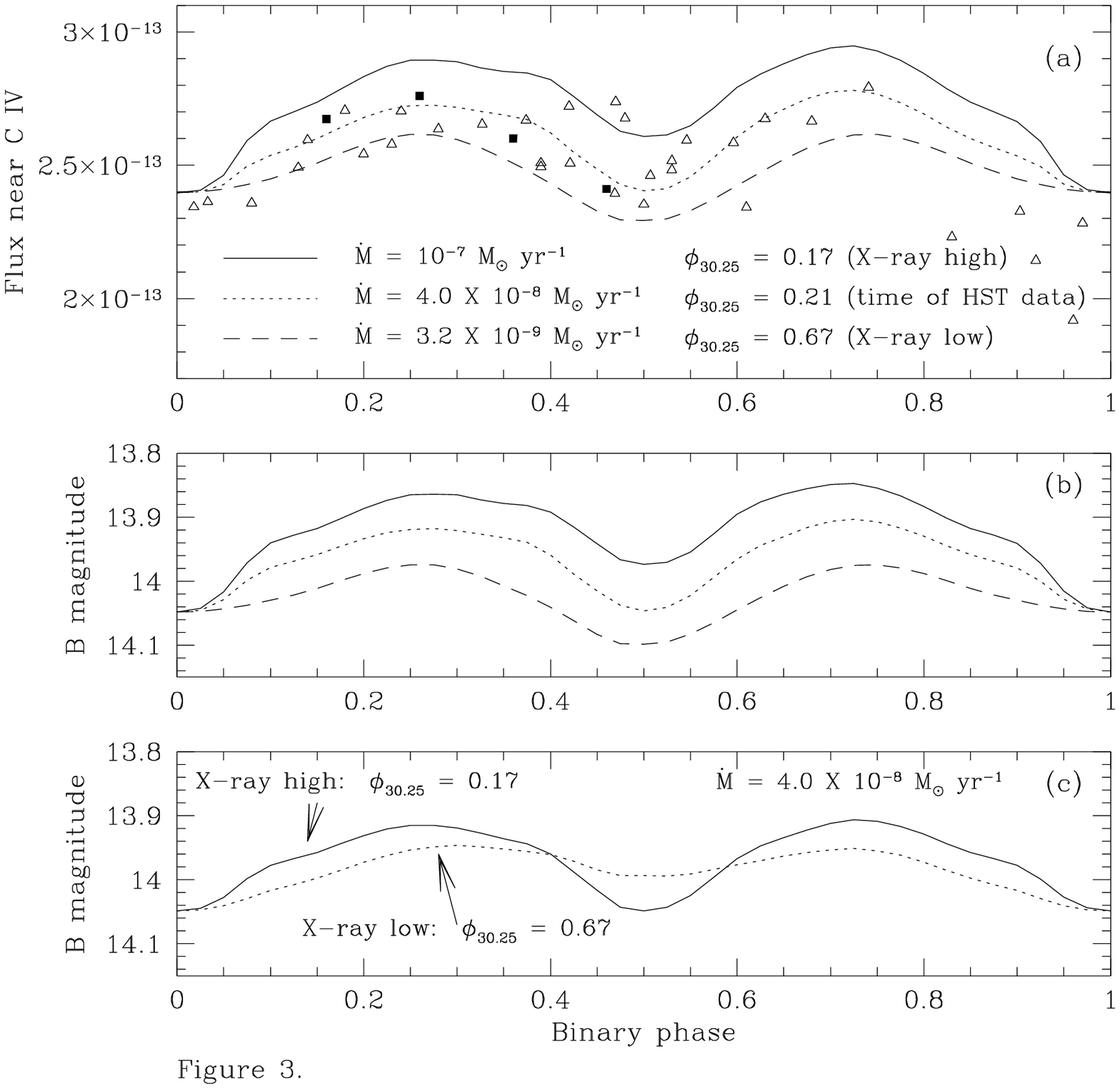}}

\newpage

\centerline{\psfig{figure=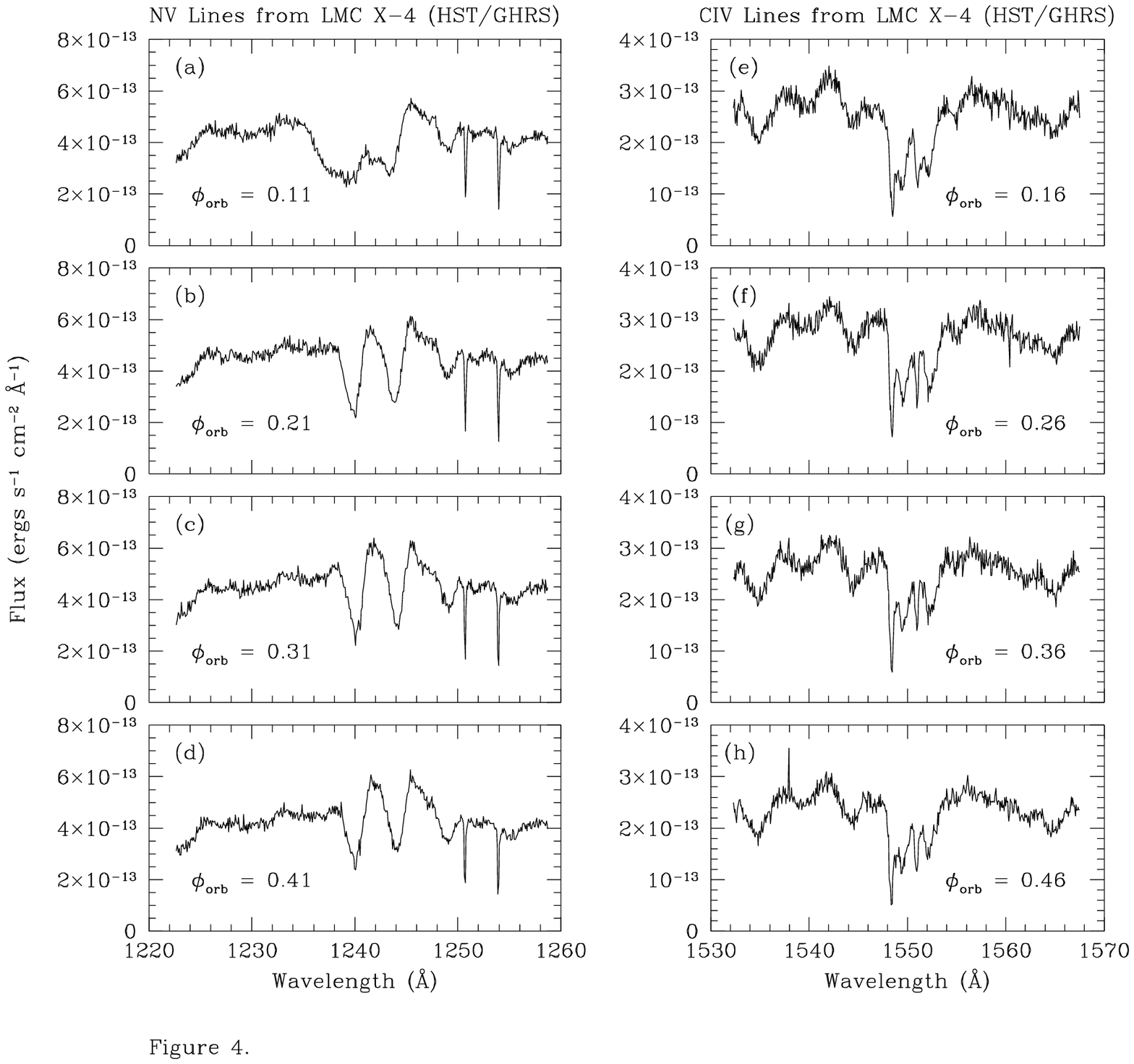}}

\newpage

\centerline{\psfig{figure=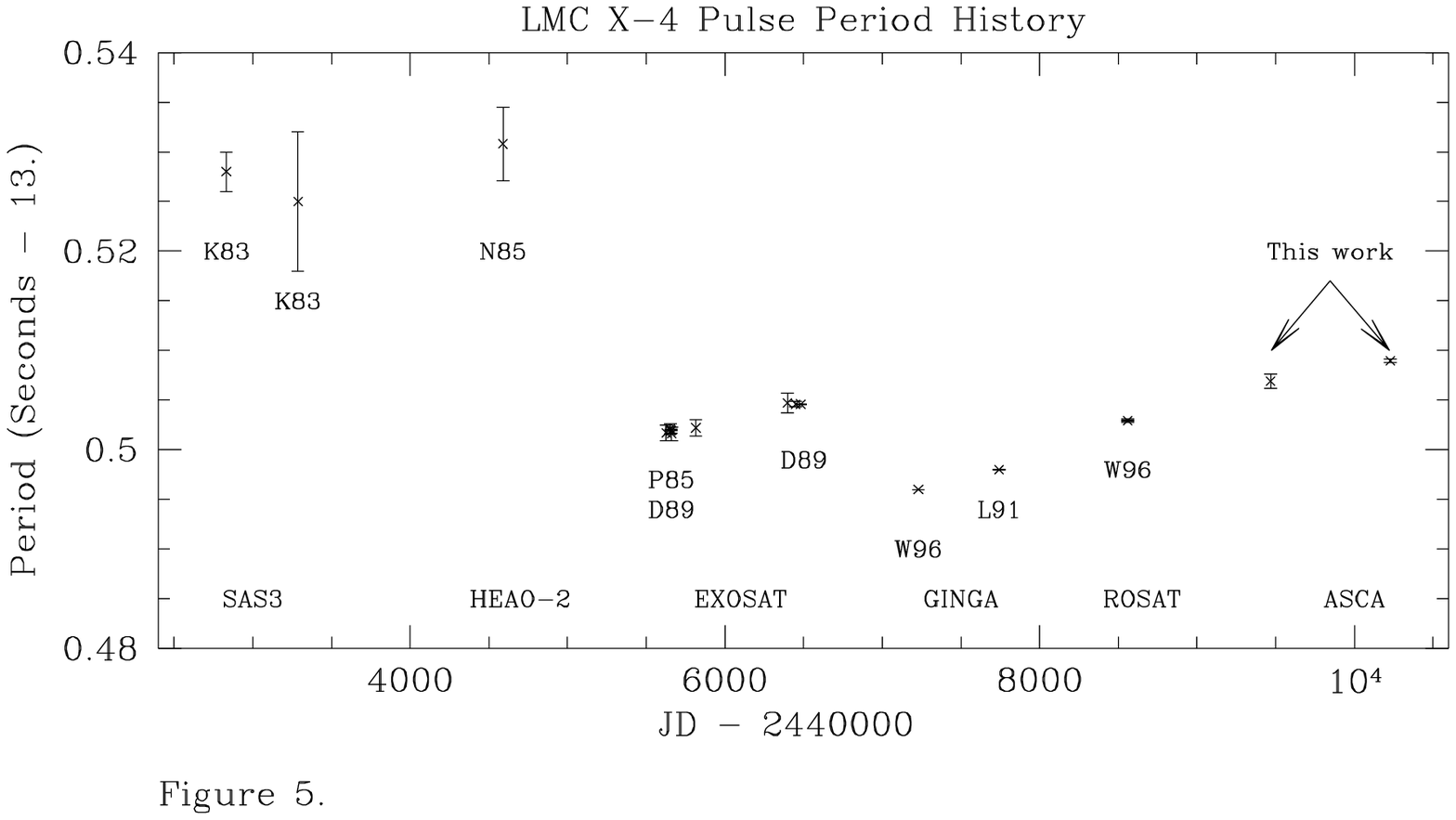}}

\end{document}